%% file: main.tex
\documentclass[prf]{revtex4-2}

\usepackage{breqn}
\usepackage{amsmath}
\usepackage{physics}
\usepackage{graphicx}
\usepackage{subcaption}
\usepackage{caption}
\usepackage{color}
\usepackage{algorithm}
\usepackage{algpseudocode}
\usepackage{mathrsfs}
\usepackage{lineno}

\makeatletter
\let\cat@comma@active\@empty
\makeatother

\newcommand{\re}{\mathrm{Re}}
\newcommand{\frechetdv}[2]{\dfrac{\mathscr{D}#1}{\mathscr{D}#2}}
\newcommand{\wallquant}[1]{\left.#1\right|_w}

\begin{document}

\title{
Bayesian optimization of a cost function in suboptimal control for reducing skin friction drag in wall turbulence
}

\date{\today}

\author{Yusuke Yugeta}
\affiliation{Department of Mechanical Engineering, The University of Tokyo, 7-3-1 Hongo, Bunkyo-Ku, Tokyo 113-8656, Japan}

\author{Yosuke Hasegawa}
\email{ysk@iis.u-tokyo.ac.jp}
\affiliation{Institute of Industrial Science, The University of Tokyo, 4-6-1 Komaba, Meguro-Ku, Tokyo 153-8505, Japan}

\input{Abstract}

\maketitle

\input{Intro}
\input{Method}

\input{SOC}

\input{BO}

\input{Results}

\input{Conclusion}

\section*{Acknowledgement}
The authors gratefully acknowledge the support by JSPS KAKENHI Grant Numbers JP23K26034 and JP21H05007.
YY acknowledges the generous support from PhD fellowship by JST Boost NAIS, Next-generation AI for Intelligent Society.
This research used the FUJITSU Supercomputer PRIMEHPC FX1000 and FUJITSU Server PRIMERGY GX2570 (Wisteria/BDEC-01) at the Information Technology Center, The University of Tokyo.

\appendix
\input{Appendix}

\bibliographystyle{apsrev4-2}
\bibliography{journal_paper_2023b}

\end{document}

%% file: Abstract.tex
\begin{abstract}

A systematic and automated framework for developing closed-loop flow control strategies is proposed, integrating suboptimal control theory [Lee \textit{et al}., J. Fluid Mech. \textbf{358}, 245 (1998)] with Bayesian optimization. 
The approach is demonstrated in the context of reducing skin friction drag in a low-Reynolds-number turbulent channel flow. 
A cost function in the suboptimal control framework is formulated as a linear combination of various wall quantities, with the corresponding weight coefficients optimized via Bayesian optimization to maximize a drag reduction rate. 
The proposed method successfully identifies effective cost functions, achieving approximately 20\% drag reduction, which is comparable to or even higher than those reported in previous studies. 
Additionally, some cost functions proposed in previous studies are rediscovered. 
The present approach offers novel perspectives on the decision of the cost function, highlighting its potential for advancing active control strategies in turbulent flows.

\end{abstract}

%% file: Intro.tex
\section{Introduction}
\label{sec:intro}
Active flow control (AFC) is a technique that utilizes actuators to manipulate the flow, aiming to achieve objectives such as drag reduction, heat transfer enhancement, and preventing flow separation, among others \cite{fukagata_turbulent_2024, kim_control_2003, gad-el-hak_flow_2000, brunton_closed-loop_2015}. 
Among various flow configurations and control objectives considered so far, reducing the skin friction drag in a fully developed turbulent channel flow is one of the most actively studied problems in the flow control community \cite{fukagata_turbulent_2024}. 
Existing active flow control algorithms can generally be categorized into predetermined and feedback controls. 
The former, typified by spanwise wall oscillation \cite{quadrio_critical_2004}, uniform wall blowing \cite{kametani_direct_2011} and a traveling wave of wall blowing and suction \cite{min_sustained_2006,fukagata_turbulent_2024}, applies a prescribed control input which is often periodic in space and/or time regardless of a flow state. 
Despite their simplicity, it is known that considerable drag reduction can be achieved. 
Meanwhile, they require relatively large power inputs for applying the controls, and therefore, the resulting net energy savings are not always satisfactory \cite{kasagi_toward_2009}. 
In contrast, a feedback control approach applies a control input that depends on the instantaneous flow state, allowing for more flexible and effective flow control. 
Meanwhile, a considerable number of sensors and actuators are generally required to detect the instantaneous flow state and apply the corresponding control input. 
Indeed, most existing studies assume that sensors and actuators cover the entire two-dimensional plane or three-dimensional volume (e.g., Ref. \cite{choi_active_1994, lee_application_1997, lee_suboptimal_1998}). 
This inevitably yields the fundamental challenges of developing a control algorithm that relates the sensing information to the complex control input. 

One of the pioneering studies in AFC is a so-called opposition control \cite{choi_active_1994}, where a zero-net-mass-flux wall blowing and suction is applied to oppose the wall-normal velocity fluctuation on a detection plane located at a certain distance from the wall. 
Specifically, when the detection plane is located at $y^+_d=10$--$15$ from the wall in the wall unit denoted by the $+$ superscript, a drag reduction rate of around $25\%$ can be obtained. 
The control algorithm is considered to disrupt or suppress the self-sustaining mechanisms of wall turbulence \cite{hamilton_regeneration_1995,jimenez_autonomous_1999,schoppa_coherent_2002}.
While the opposition control is one of the earliest and most successful algorithms developed based on the physics of near-wall turbulence, further development of novel control algorithms with a similar approach turned out to be quite challenging due to the complexity of a turbulent flow and also the lack of our understanding of the physical phenomena.
As a result, the community has begun to explore other approaches simultaneously, such as model-based and black-box (or adaptive) methods.

As an example of model-based optimization, optimal control theory, first applied to a flow control problem by Abergel and Temam \cite{abergel_control_1990}, is one of the most potent methods for optimizing a control input based on flow physics mathematically. 
By applying the optimal control theory, Bewley et al.\cite{bewley_dns-based_2001} achieved the relaminarization of a turbulent channel flow at a low Reynolds number. 
In previous studies applying the optimal control theory \cite{bewley_dns-based_2001,choi_feedback_1993,li_optimal_2003,yamamoto_optimal_2013,mao_nonlinear_2015,yugeta_prediction_2023}, the spatial distributions of control inputs with approximately $10^4$ or even more degrees of freedom and also their time evolutions have been successfully optimized by explicitly considering the governing equations of the flow, i.e., the Navier-Stokes and mass conservation equations. One shortcoming of the optimal control is that it requires expensive forward-adjoint iterations to find the optimal control input within a prescribed time horizon in which a cost function to be minimized is defined. 
In addition, the complete space-time information of the flow field needs to be stored for the adjoint analysis. 
To overcome the difficulties, the suboptimal control theory, where an optimal control input for a vanishingly small time horizon is considered, has been developed and applied to various flow control problems \cite{lee_suboptimal_1998,fukagata_suboptimal_2004,hasegawa_dissimilar_2011,iwamoto_reynolds_2002,min_suboptimal_1999,naito_control_2014,choi_assessment_2002,nakashima_assessment_2017,tardu_one-information_2009,kawagoe_proposal_2019,kang_suboptimal_2002,jeon_suboptimal_2010,xu_suboptimal_2002}. 
It has been demonstrated that the suboptimal control can reduce drag comparable to or even more than the opposition control despite using wall information only. 
Although applying the optimal control theory to flow control problems has great potential as mentioned above, there remain obstacles to be resolved.

In optimal control theory, the control input is deterministically obtained so as to satisfy the stationary condition for minimizing a given cost function based on its gradient. Although a long-term objective of control is clearly stated, setting it directly as a cost function does not guarantee the success of the optimal control. This is attributed to the fact that the time horizon in which the control input is optimized is always finite, and much shorter than the time scale over which the flow state reaches another equilibrium state under the applied control. 
More specifically, in the case of wall turbulence, the maximum length of the time horizon used in the optimal control is around 100 in the wall unit, while the time period required for the flow to reach the equilibrium state after applying control and for flow statistics to converge is at least 10--100 times longer.

Bewley et al. \cite{bewley_dns-based_2001} compared three cost functions, namely, the time-averaged streamwise shear stress within the time horizon, turbulent kinetic energy (TKE) averaged within the time horizon, and TKE at the end of the time horizon.
Interestingly, while the last one achieves re-laminarization corresponding to around 60\% drag-reduction rate in their configuration, the drag reduction effects achieved by the others are limited to at most 20-25\% comparable to that achieved by the opposition control. 
As for the suboptimal control, Lee et al. \cite{lee_suboptimal_1998} examined three cost functions, namely, the fluctuations of the streamwise and spanwise shear stresses and also the pressure gradient in the spanwise direction. The three cost functions result in quite different drag reduction rates of 0\%, 25\%, and 16\%, respectively.
Inspired by the so-called Fukagata-Iwamoto-Kasagi identity \cite{fukagata_contribution_2002,kasagi_control_2012} relating the wall-normal distribution of the Reynolds shear stress and the friction coefficient $C_f$, subsequent studies \cite{fukagata_suboptimal_2004,hasegawa_dissimilar_2011} have included the Reynolds shear stress to the cost function in the suboptimal control framework. Even in these cases, however, 
the obtained drag reduction rates are similar to 
that achieved by Lee et al. \cite{lee_suboptimal_1998}.
In summary, although the choice of the cost function has significant impacts on the resulting control performance, no systematic methodology for finding effective cost functions has been established yet.

Recently, black-box optimization (BBO) or adaptive optimization approaches, such as genetic algorithms (GA) \cite{yoshino_drag_2008}, Bayesian optimization (BO) \cite{mahfoze_reducing_2019,blanchard_bayesian_2021,nabae_bayesian_2021,mallor_bayesian_2023,pino_comparative_2023,oconnor_optimisation_2023} and reinforcement learning (RFL) \cite{rabault_artificial_2019,fan_reinforcement_2020,paris_robust_2021,sonoda_reinforcement_2023,vignon_recent_2023,lee_turbulence_2023,pino_comparative_2023,guastoni_deep_2023}, have achieved promising results in AFC and general flow optimization problems (see Ref. \cite{morita_applying_2022,wang_multi-objective_2023}, for example).
Although such black-box optimizations (BBO) are flexible in their applications, there remain two key issues when they are applied to flow control problems, especially involving turbulent flows. 
The first issue is that a typical number of degrees of freedom of a control input handled in BBO is limited to around $O(10)$, while the degrees of freedom of a turbulent flow and a control input is generally much larger. 
Due to this difficulty, most previous studies considered control inputs with only a few numbers of degrees of freedom, such as two jets for controlling a two-dimensional flow around a cylinder (e.g., Ref. \cite{rabault_artificial_2019}), and a few control points to describe the spatial distribution of the control input (e.g., Ref. \cite{mallor_bayesian_2023,wang_multi-objective_2023,morita_applying_2022,oconnor_optimisation_2023}). 
There also exist other approaches to effectively reduce the degrees of freedom of the control input by leveraging the statistical homogeneity of the targeted flow systems
\cite{sonoda_reinforcement_2023,guastoni_deep_2023,lee_turbulence_2023,peitz_distributed_2024}
or introducing surrogate models (e.g., Ref. \cite{bagheri_inputoutput_2009,kim_control_2003,mohan_deep_2018,noack_low-dimensional_2004,bergmann_optimal_2005,linot_turbulence_2023}).
The second issue in BBO is that the mathematical models of considered flow systems such as the Navier-Stokes and continuity equations are not explicitly incorporated into the optimization procedure, and their features may be learned only through numerous trials with different control inputs. 
This contrasts with the optimal control theory, where the mathematical models are explicitly integrated as constraints in the optimization procedure.
Although BBO approaches have the potential to find novel and effective control strategies, which are difficult for humans to conceive otherwise (e.g., Sonoda et al. \cite{sonoda_reinforcement_2023}), the above two issues have to be resolved to extend the applicability of BBO to flow problems.

The objective of the present study is to propose a new optimization approach to compensate for the above-mentioned shortcomings of the model-based and black-box optimizations and leverage their advantages. 
Specifically, we formulate the cost function as a weighted sum of various wall quantities. 
By applying the suboptimal control theory \cite{lee_suboptimal_1998}, the control input can be analytically derived for the newly proposed cost function with arbitrary weight coefficients for the wall quantities. 
Meanwhile, BBO optimizes the weight of each wall quantity in the cost function through trial and error. 
This way, we can enjoy the high adaptivity of BBO, while maintaining the unique strength of the optimal control theory that it enables to optimize control variables with large degrees of freedom by explicitly taking into account the mathematical model of a considered system.

The rest of the manuscript proceeds as follows:
In Sec.~\ref{method}, the present flow configuration and numerical schemes for solving the flow field are introduced. Then, the formulation of the suboptimal control is explained in Sec.~\ref{SOC}. In section \ref{BO}, Bayesian optimization frameworks and metrics to be minimized are introduced, whereas
the obtained control performances and their interpretations are discussed in Sec.~\ref{results}. Finally, the present study is summarized in Sec.~\ref{conclusion}.

%% file: Method.tex
\section{Problem setups and numerical configurations}
\label{method}
\begin{figure}[h]
    \centering
    \includegraphics[width=0.5\hsize]{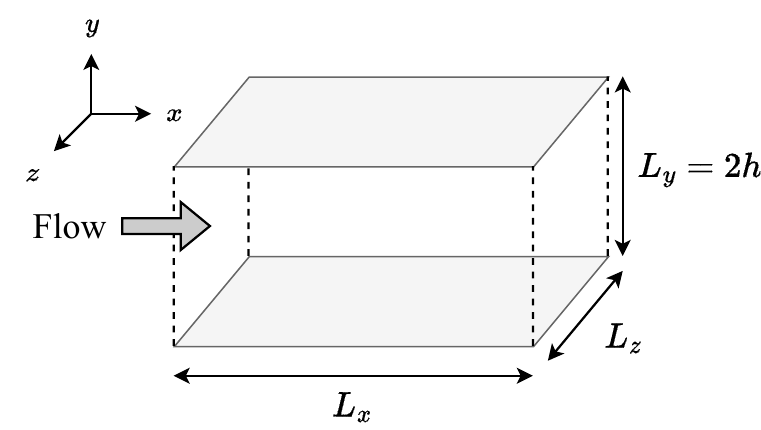}
    \caption{The computational domain and coordinate system.}
    \label{fig:computational-domain}
\end{figure}

We consider a fully developed turbulent channel flow under a constant flow rate condition as schematically presented in Fig. \ref{fig:computational-domain}.
The streamwise, wall-normal, and spanwise directions are denoted by $(x,y,z)$, while their corresponding velocity components are $(u,v,w)$, respectively. 
The time and static pressure are represented by $t$ and $p$, respectively.
For ease of notation, we also denote $(x,y,z)$ with $x_i$, whereas $(u,v,w)$ with $u_i$, where $i$ is a free index of a vector. 
The bottom and top walls are placed at $y=0$ and $y=2$, respectively. 
Periodic boundary conditions are applied in the streamwise and spanwise directions. 
In the present study, we consider wall blowing and suction as a control input. 
Therefore, no-slip conditions are imposed on the top and bottom walls only for the streamwise and spanwise velocity components, $u_1$ and $u_3$, whereas the wall-normal velocity $u_2$ on the wall is determined based on the suboptimal control theory as will be explained in detail in the next section. 
The fluid is assumed to be incompressible and Newtonian, and therefore its dynamics can be described by the following momentum and mass conservations:

\begin{equation}
    \pdv{u_i}{t} + \pdv{(u_iu_j)}{x_j} = -\pdv{p}{x_i} + \dfrac{1}{\re_b}\pdv{u_i}{u_j}{u_j},
    \label{eq:NS}
\end{equation}

\begin{equation}
    \pdv{u_i}{x_i} = 0,
    \label{eq:Con}
\end{equation}

\noindent
where all the variables are normalized using the bulk mean velocity $U_b^\dagger$ and the channel half-height $h^\dagger$. 
The superscript of a dagger $\bullet^\dagger$ indicates a dimensional quantity.
In the present study, the flow rate $Q$ is kept constant regardless of the presence of the control. 
Accordingly, the bulk Reynolds number is set to be $\re_b=2U_b^\dagger h^\dagger/\nu^\dagger=3220$, which corresponds to the friction Reynolds number of $\re_\tau=u_\tau^\dagger h^\dagger/\nu^\dagger\approx110$ for the uncontrolled flow. 
Here, $\nu^\dagger$ is a kinematic viscosity, and $u_\tau^\dagger$ is a friction velocity defined as  $u_\tau^\dagger\equiv\sqrt{\nu^\dagger\overline{(\partial{u^\dagger}/\partial{y^\dagger})_{w}}}$. The subscript of $w$ represents a wall quantity.
Unless otherwise noted, we indicate the wall unit with the superscript $\bullet^+$ and define it using the friction velocity of the uncontrolled flow.
$\overline{\bullet}$ indicates the average in the homogeneous directions of $x$ and $z$ as well as time $t$.
The dimensions of the computational domain in the streamwise, wall-normal and spanwise directions are set to be $(L_x, L_y, L_z) = (4\pi, 2.0, 4\pi/3)$.

To solve the governing equations (\ref{eq:NS}-\ref{eq:Con}), we adopt direct numerical simulation (DNS) based on an open-source code, Incompact3d \cite{laizet_high-order_2009,laizet_incompact3d_2011}. 
The solver utilizes a Cartesian mesh uniformly distributed in the streamwise and spanwise directions with its wall-normal grid spacing being gradually stretched as increasing the distance from the wall
\cite{laizet_incompact3d_2011}. 
For spatial discretization, the six-order compact scheme \cite{laizet_high-order_2009} is employed. 
As for the time integration, the second-order Crank-Nicolson method is adopted for the viscous terms in the wall-normal direction (i.e., ${\partial^2}/{\partial y^2}$), whereas the third-order Adams-Bashforth method is applied for the other terms. 
Equations~\eqref{eq:NS} and \eqref{eq:Con} are coupled through a fractional step method\cite{kim_application_1985}.
The pressure Poisson equation 
is directly solved by the use of a generic discrete three-dimensional Fourier transformation to avoid cumbersome iterative computation\cite{laizet_incompact3d_2011}. 
For further details about the numerical solver, readers are referred to Ref. \cite{laizet_high-order_2009,laizet_incompact3d_2011} and their website (\url{www.incompact3d.com}).

\begin{figure}[h]
    \centering
    \begin{minipage}[h]{0.40\hsize}
        \centering
        \includegraphics[width=\hsize]{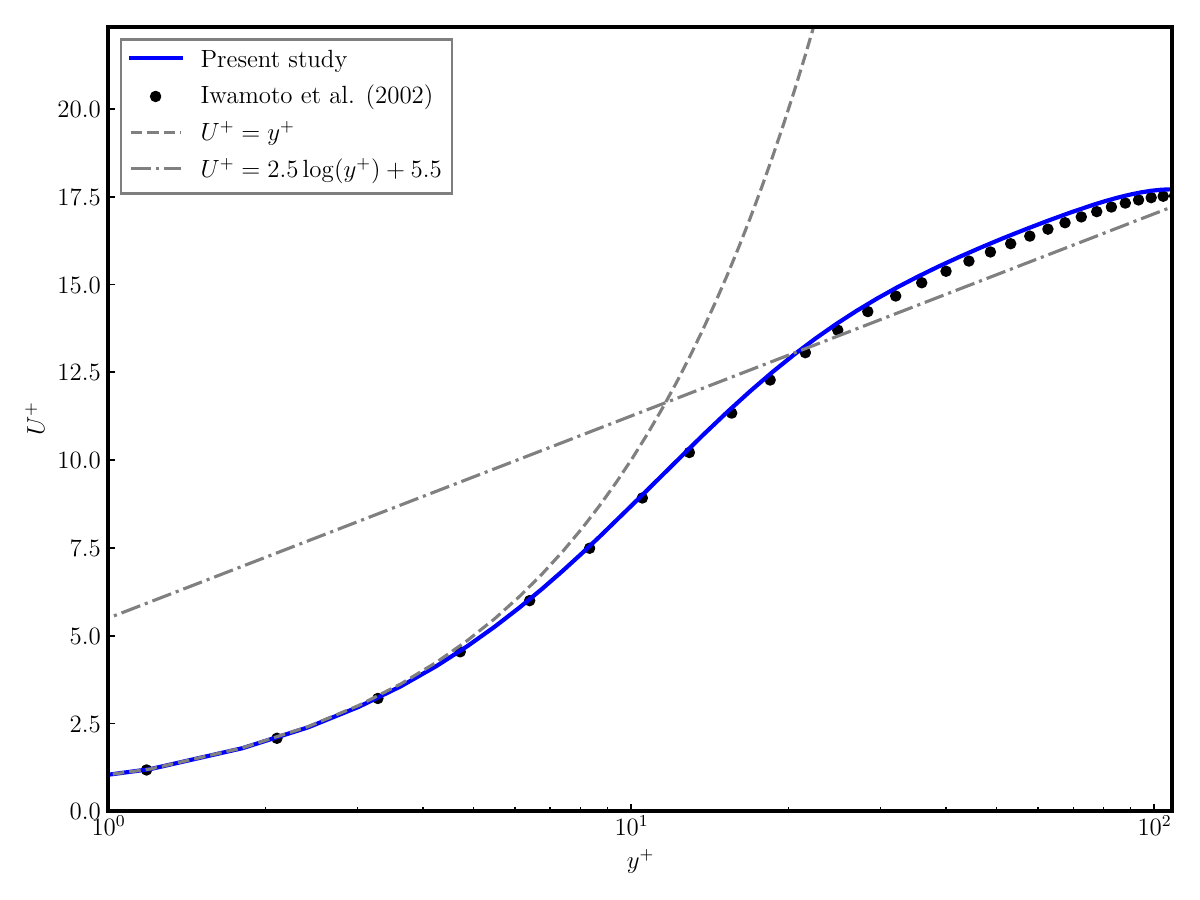}
        \subcaption{Mean velocity profile}
        \label{fig:mean-velocity}
    \end{minipage}
    \begin{minipage}[h]{0.40\hsize}
        \centering
        \includegraphics[width=\hsize]{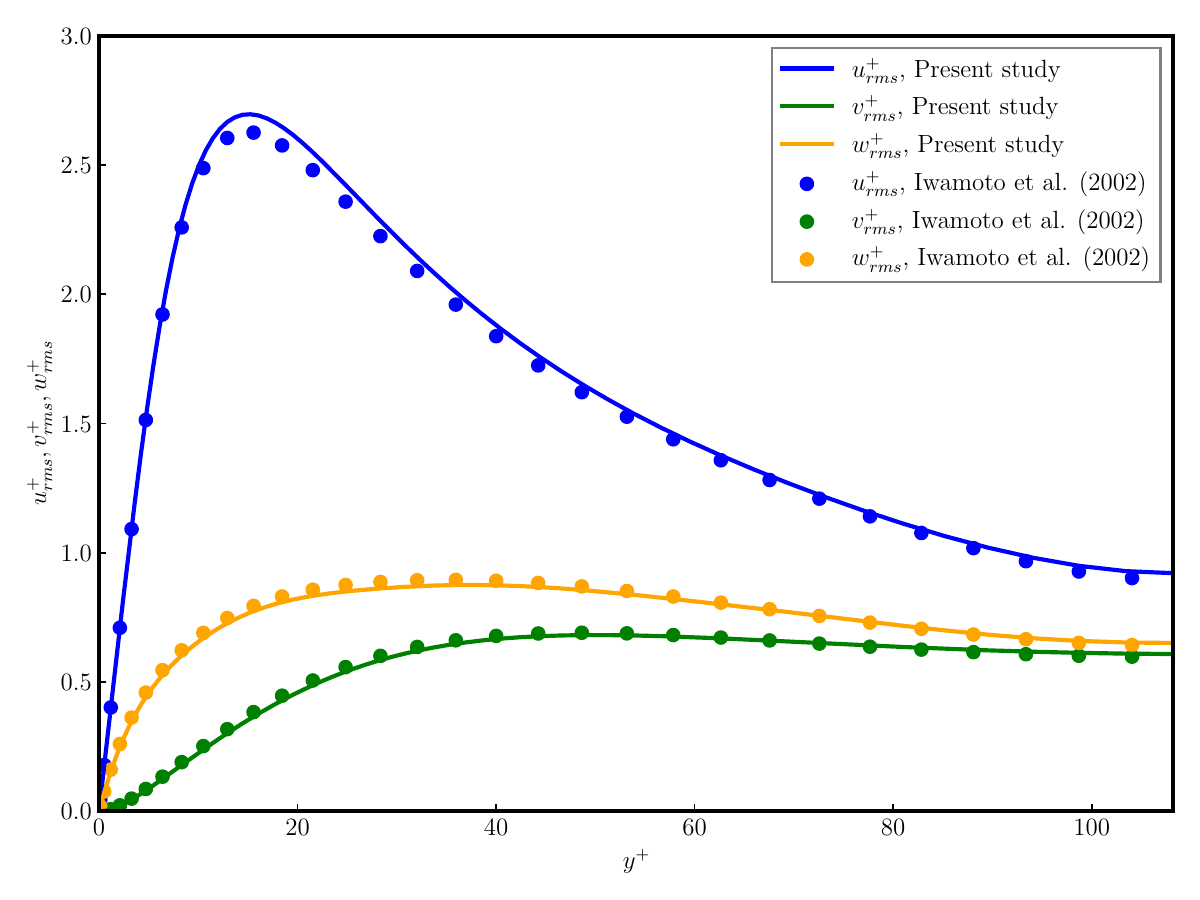}
        \subcaption{RMS profile}
        \label{fig:reynolds-stress}
    \end{minipage}
    \caption{The mean velocity and RMS profiles of the uncontrolled flow. The solid lines are the present results, whereas the markers are the  reference\cite{iwamoto_reynolds_2002}.}
    \label{fig:validation}
\end{figure}

The numbers of grid points in the three directions are set as $(N_x, N_y, N_z) = (128, 129, 96)$.
This setup corresponds to the grid spacings of $(\Delta x^+, \Delta y^+, \Delta z^+, \Delta t^+)=(10.7,0.6$--$4.8,4.8,0.02)$.
As shown in Fig. \ref{fig:validation}, we have verified that the present simulation reproduces basic statistics of the uncontrolled flow, such as the mean velocity profile and the root-mean-square values of the velocity fluctuations reported in a previous study \cite{iwamoto_reynolds_2002}. 
As for flows with controls, we have tested several cost functions proposed in previous studies \cite{choi_active_1994,lee_suboptimal_1998,iwamoto_reynolds_2002,choi_assessment_2002}, and confirmed that the obtained drag reduction rates agree reasonably well with those reported in the literature (not shown here). 

%% file: SOC.tex
\section{Suboptimal Control}
\label{SOC}
\subsection{cost function}
We apply the suboptimal control theory \cite{lee_suboptimal_1998} to determine the control input, i.e., wall blowing and suction. 
As mentioned in the Introduction, the suboptimal control considers a short time horizon so that the spatio-temporal distribution of wall blowing and suction is determined to minimize a prescribed cost function $\mathcal{J}$ at the next computational time step.
Since we consider a vanishingly small time horizon, the choice of the cost function is not trivial for achieving a given long-term objective, i.e., drag reduction in the present study.
In the present study, we consider a linear weighted sum of various wall quantities as a cost function as follows:

\begin{eqnarray}
    \mathcal{J} \equiv \left\langle \phi^2 + a_1 \left. \pdv{u}{y}\right|_w^2 + a_2 \left.\pdv{u}{y}\right|_w\left.\pdv{w}{y}\right|_w + a_3 \left.\pdv{u}{y}\right|_w p_w + a_4 \left. \pdv{u}{y}\right|_w\left.\pdv{p}{x}\right|_w + a_5 \left. \pdv{u}{y}\right|_w\left.\pdv{p}{z}\right|_w + a_6 \left. \pdv{w}{y}\right|_w^2 + \right. \nonumber\\
    \left.
    a_7 \left.\pdv{w}{y}\right|_w p_w + a_8 \left. \pdv{w}{y}\right|_w\left.\pdv{p}{x}\right|_w + a_9 \left. \pdv{w}{y}\right|_w\left.\pdv{p}{z}\right|_w + a_{10} p_w^2 + a_{11} \left.\pdv{p}{x}\right|_w^2 + a_{12} \left.\pdv{p}{x}\right|_w\left.\pdv{p}{z}\right|_w + a_{13} \left. \pdv{p}{z} \right|_w^2
    \right\rangle,
    \label{eq:cost}
\end{eqnarray}
\noindent 
where the bracket is defined as:
\begin{equation}
    \langle\bullet\rangle\equiv 
    \frac{1}{L_xL_z\Delta t}\int_t^{t+\Delta t}\int_{0}^{L_x}\int_{0}^{L_z}\bullet~dzdxdt
\end{equation}
The time horizon is set to be $\Delta t^+ = 1.0$ in the present study, and its value is sufficiently shorter than a typical turbulence time-scale near the wall. 
The weight coefficients $a_i~(i=1,2,\dots,13)$ are constants to be optimized for achieving a better control performance, i.e., a drag reduction rate after the controlled flow reaches a statistically equilibrium state.
As will be derived later, since the present cost function includes wall quantities only, the resulting suboptimal control input is analytically obtained as a convolution of these wall values. 
To ensure the net mass flux is always zero, we neglect the contributions from the Fourier mode with $(k_x,k_z)=(0,0)$ of the wall quantities, where $k_x$ and $k_z$ are the streamwise and spanwise wavenumbers.
 
In Eq. \eqref{eq:cost}, we exclude two products of wall quantities such as $p{\partial p}/{\partial x}$ and $p{\partial p}/{\partial z}$ from the cost function. 
This is because they can be written as 
$\left\langle p{\partial p}/{\partial x}\right\rangle =\left\langle{1}/{2}{\partial p^2}/{\partial x}\right\rangle=0$ and $\left\langle p{\partial p}/{\partial z}\right\rangle =\left\langle{1}/{2}{\partial p^2}/{\partial z}\right\rangle=0$, and therefore they are exactly zero due to the periodicity in the $x$ and $z$ directions. 
In addition, some cross-correlation such as $\left.{\partial u}/{\partial y}\right|_w\left.{\partial w}/{\partial y}\right|_w$ are zero for the uncontrolled flow when the time integral is taken sufficiently long time due to the spanwise symmetry of the present flow configuration.  
In the controlled flow, however, the spanwise symmetry is not always guaranteed due to possible asymmetry of the applied control input, and also the short time horizon. 
Therefore, these terms are kept in the cost function \eqref{eq:cost}.

In the present cost function \eqref{eq:cost}, the constants $a_i~(i=1,2,\dots,13)$ represent the importance of each term relative to the cost for applying the control input, i.e., the first term $\phi^2$ inside the bracket on the right-hand-side. 
However, each wall quantity appearing in the cost function \eqref{eq:cost} could have a different magnitude so that the values of the constants $a_1 - a_{13}$ may not simply correspond to the relative importance of each term, but may depend on the magnitude of each wall quantity. 
Therefore, we introduce additional coefficients $w_i\in[-1,1]$, and they are related to $a_i$ as follows:
\begin{equation}
    a_i = C(t)\dfrac{w_i}{\phi^{\mathrm{RMS}}_i(t=0)}.
    \label{eq:weight_factor}
\end{equation}
Here, $\phi_i^{\mathrm{RMS}}(t=0)$ refers to the RMS value of $\phi_i$, which is the contribution from the $i$-th term in the cost function \eqref{eq:cost} to the control input at the onset of the control ($t=0$).
This way, $w_i$ can be considered more suitable parameters to evaluate the relative importance of each term in the cost function \eqref{eq:cost}. 
In the present study, $w_i$ are optimized through the Bayesian optimization. 
The factor $C(t)$ common to all the weights is adjusted at each time step to make the RMS value of the control input the same as that of the instantaneous wall-normal velocity fluctuation at $y^+_d=11.7$. 
The normalization of the magnitude of the control input is introduced to fairly compare the present scheme with the counterpart, i.e., opposition control \cite{choi_active_1994}.
In the present study, we maintain the magnitude of the control input at $\phi_{rms}^+\sim0.1$.

In this paper, we conducted two series of optimization campaigns, i.e., Case 1 and Case 2.
Case 1 considers the cost function \eqref{eq:cost_Case 1}, which has two terms, while Case 2 considers the cost function \eqref{eq:cost_Case 2} with ten terms.
According to the reference \cite{lee_suboptimal_1998}, the second term $({\partial u}/{\partial y})^2$ in the cost function \eqref{eq:cost_Case 1} does not have any control effect, whilst the third term $({\partial w}/{\partial y})^2$ shows a substantial drag reduction rate when $a_6<0$.
Case 1 is to confirm that the present framework can find the existing control strategy under a simple scenario, while Case 2 is to explore a new control strategy.
In Case 2, we note that we always fix $a_3,a_7,~\mathrm{and~}a_{10}$ as zero due to numerical instability.
We also note that the resultant drag reduction rate was comparable to or less than that of the existing one and the cases presented in the manuscript, even when the pressure is included (results for the pressure-included case are not shown here).
\begin{equation}
    \mathcal{J} \equiv \left\langle\phi^2+a_1\wallquant{\pdv{u}{y}}^2+a_6\wallquant{\pdv{w}{y}}^2\right\rangle
    \label{eq:cost_Case 1}
\end{equation}
\begin{eqnarray}
    \mathcal{J} \equiv \left\langle \phi^2 + a_1 \left. \pdv{u}{y}\right|_w^2 + a_2 \left.\pdv{u}{y}\right|_w\left.\pdv{w}{y}\right|_w + a_4 \left. \pdv{u}{y}\right|_w\left.\pdv{p}{x}\right|_w + a_5 \left. \pdv{u}{y}\right|_w\left.\pdv{p}{z}\right|_w + \right.\nonumber\\
    \left.a_6 \left. \pdv{w}{y}\right|_w^2 + a_8 \left. \pdv{w}{y}\right|_w\left.\pdv{p}{x}\right|_w + a_9 \left. \pdv{w}{y}\right|_w\left.\pdv{p}{z}\right|_w + a_{11} \left.\pdv{p}{x}\right|_w^2 + a_{12} \left.\pdv{p}{x}\right|_w\left.\pdv{p}{z}\right|_w + a_{13} \left. \pdv{p}{z} \right|_w^2
    \right\rangle.
    \label{eq:cost_Case 2}
\end{eqnarray}

\subsection{Derivation of the suboptimal control input}

\subsubsection{Response of the velocity field to wall blowing and suction}

We optimize the control input to minimize the cost function \eqref{eq:cost} for a sufficiently short time horizon so that only rapid linear processes are considered, while non-linear processes are omitted. 
Accordingly, the response of the velocity field at the time step of $n+1$ subject to wall blowing and suction $u_2^n(x,0,z)=\phi^n(x,z)$ applied at the time step of $n$ can be approximated by the following equations:
\begin{equation}
    u_i^{n+1}-\frac{\Delta t}{2\re}\pdv{u_i^{n+1}}{x_j}{x_j}+\frac{\Delta{t}}{2}\pdv{p^{n+1}}{x_i}=R_i^n.
    ~\mathrm{and}~
    \pdv{u_i^{n+1}}{x_j}=0.
    \label{eq:lin-NS}
\end{equation}
We note that Eqs. \eqref{eq:lin-NS} are written in a discretized form in time based on the Crank-Nicolson method for linear terms and the Euler-forward method for non-linear terms. 
$\Delta t$ is a computational time step. 
$R_i^n$ is the residual terms in the original Navier-Stokes equations \eqref{eq:NS} including the non-linear terms and the part of pressure gradient and viscous terms which are treated explicitly and thereby only depend on the flow field at the time step of $n$.

We then define the Fr\'{e}chet derivatives $\mathscr{D}$ of the velocity and pressure fields at the next time step $n+1$ variables concerning the control input $\phi$ as

\begin{equation}
    \theta_i \equiv \frechetdv{u_i}{\phi}\tilde{\phi} \equiv \lim_{\epsilon\rightarrow0}\frac{u_i(\phi+\epsilon \tilde{\phi})-u_i(\phi)}{\epsilon},~\mathrm{and}~\sigma \equiv \frechetdv{p}{\phi}\tilde{\phi},
\end{equation}

\noindent where $\tilde{\phi}$ is a perturbation in the control input and $\epsilon$ is an infinitesimal value.

Applying the Fr\'echet derivatives to Eqs. \eqref{eq:lin-NS}, the equations for the differential states of the velocity and pressure fields are obtained as

\begin{equation}
    \theta_i^{n+1}-\frac{\Delta t}{2\re}\pdv{\theta_i^{n+1}}{x_j}{x_j}+\frac{\Delta{t}}{2}\pdv{\sigma_i^{n+1}}{x_i}=0,
    ~\mathrm{and}~
    \pdv{\theta_i^{n+1}}{x_i} = 0.
    \label{eq:lin-frechet-NS}
\end{equation}

The solutions to the above equations near the bottom wall at $y=0$ can be derived as
\begin{equation}
\begin{array}{c}
    \widehat{\theta_1}(k_x,y,k_z)=\dfrac{ik_x}{K}\widehat{\tilde{\phi}}\left(\exp{(-(2\re/\Delta t)^{1/2}y)}-\exp{(-Ky)}\right),\\
    \widehat{\theta_2}(k_x,y,k_z)=\widehat{\tilde{\phi}}\exp{(-Ky)},\\
    \widehat{\theta_3}(k_x,y,k_z)=\dfrac{ik_z}{K}\widehat{\tilde{\phi}}\left(\exp{(-(2\re/\Delta t)^{1/2}y)}-\exp{(-Ky)}\right),\\
    \widehat{\sigma}(k_x,y,k_z) = \dfrac{2}{k\Delta t}\widehat{{\tilde{\phi}}}\exp{(-Ky)}.
    \label{eq:approx-sol}
\end{array}
\end{equation}

Here, considering the periodicity in the streamwise and spanwise directions, we introduce the Fourier mode as $\theta_i(y)\equiv\sum_m\sum_n\widehat{\theta_i(y)}\exp{(i(k_xx+k_zz))}$, where the streamwise and spanwise wavenumbers and their norm are defined as $k_x\equiv2\pi m/L_x$, $k_z\equiv2\pi n/L_z$, and $K=\sqrt{k_x^2+k_z^2}$, respectively. 
The above solutions are derived assuming that $\sqrt{{2\re}/{\Delta t}}$ is sufficiently larger than $K$, and the solutions are finite when $y\rightarrow\infty$ (see, the Appendix of Ref. \cite{lee_suboptimal_1998} for more detailed discussions) in Eqs. \eqref{eq:approx-sol}. The response of the velocity field to the applied control input at the top wall can also be derived similarly.

\subsubsection{Stationary condition of the cost function}
\label{subsubsec:stationaryondition}
Since the contribution from each term in the cost function \eqref{eq:cost} to the suboptimal control input is linear, each term can be treated independently, and the suboptimal control input can be given as a linear sum of each contribution with the weight coefficient $a_i$. Due to the above reason and space limitation, we showcase the derivation of the suboptimal control input considering only $\phi_2$ and $\phi_4$ terms below.
Then, the cost function reduces to 

\begin{equation}
    \mathcal{J}_e \equiv \left\langle \phi^2 + a_2\wallquant{\pdv{u}{y}}\wallquant{\pdv{w}{y}} + a_4\wallquant{\pdv{u}{y}}\wallquant{\pdv{p}{x}} \right\rangle.
\end{equation}

Applying the Fr\'{e}chet derivative to the above equation leads to

\begin{equation}
    \frechetdv{\mathcal{J}_e}{\phi}\tilde{\phi} = \left\langle 2\phi\tilde{\phi} 
    + a_2 \wallquant{\pdv{\theta_1}{y}}\wallquant{\pdv{w}{y}} + a_2 \wallquant{\pdv{u}{y}}\wallquant{\pdv{\theta_3}{y}}
    + a_4 \wallquant{\pdv{\theta_1}{y}}\wallquant{\pdv{p}{x}} + a_4 \wallquant{\pdv{u}{y}}\wallquant{\pdv{\sigma}{x}}
    \right\rangle
    \label{eq:cost_frechet}.
\end{equation}

Secondly, we consider two-dimensional discrete Fourier transformation for $x$ and $z$ direction. 
Then, the cost function \eqref{eq:cost_frechet} is rewritten as:

\begin{equation}
    \widehat{\frechetdv{\mathcal{J}_e}{\phi}}\widehat{\tilde{\phi}}^* = 2\phi\tilde{\phi} 
    + a_2 \widehat{\wallquant{\pdv{\theta_1}{y}}}^*\widehat{\wallquant{\pdv{w}{y}}} + a_2 \widehat{\wallquant{\pdv{u}{y}}}\widehat{\wallquant{\pdv{\theta_3}{y}}}^*
    + a_4 \widehat{\wallquant{\pdv{\theta_1}{y}}}^*\widehat{\wallquant{\pdv{p}{x}}} + a_4 \widehat{\wallquant{\pdv{u}{y}}}\widehat{\wallquant{\pdv{\sigma}{x}}}^*.
    \label{eq:cost_wave_1}
\end{equation}

From Eqs. \eqref{eq:approx-sol} with the assumption $\sqrt{{2\re}/{\Delta t}} \gg K$, we obtain

\begin{equation}
    \widehat{\wallquant{\pdv{\theta_{1}}{y}}}^* \approx \sqrt{\dfrac{2\re}{\Delta t}}\frac{ik_x}{K}\widehat{\tilde{\phi}}^*,~\widehat{\wallquant{\pdv{\theta_{3}}{y}}}^* \approx \sqrt{\dfrac{2\re}{\Delta t}}\frac{ik_z}{K}\widehat{\tilde{\phi}}^*,~\mathrm{and}~\widehat{\sigma}\approx\dfrac{2}{K\Delta t}\widehat{\tilde{\phi}}.
    \label{eq:lee_borrow}
\end{equation}

By substituting Eqs. \eqref{eq:lee_borrow} into Eq. \eqref{eq:cost_wave_1}, we can rearrange Eq. \eqref{eq:cost_wave_1} as

\begin{equation}
    \widehat{\frechetdv{\mathcal{J}_e}{\phi}}\widehat{\tilde{\phi}}^* = \left(\widehat{\phi}
    + a_2 \sqrt{\dfrac{2\re}{\Delta t}}\frac{ik_x}{2K} \widehat{\wallquant{\pdv{w}{y}}} + a_2 \sqrt{\dfrac{2\re}{\Delta t}}\frac{ik_z}{2K} \widehat{\wallquant{\pdv{u}{y}}}
    - a_4 \sqrt{\dfrac{2\re}{\Delta t}}\frac{k_x^2}{2K} p_w
    - a_4 \dfrac{ik_x}{K\Delta t} \widehat{\wallquant{\pdv{u}{y}}}\right)\times2\widehat{\tilde{\phi}}^*.
    \label{eq:cost_wave_2}
\end{equation}

From the stationary condition, i.e., ${\mathscr{D}\mathcal{J}}/{\mathscr{D}\phi}=0$,we obtained the following suboptimal control input:

\begin{equation}
    \widehat{\phi}(k_x,k_z) = - a_2 \left(\sqrt{\dfrac{2\re}{\Delta t}}\frac{ik_x}{2K} \widehat{\wallquant{\pdv{w}{y}}} + \sqrt{\dfrac{2\re}{\Delta t}}\frac{ik_z}{2K}\widehat{\wallquant{\pdv{u}{y}}}\right)
    + a_4 \left(\sqrt{\dfrac{2\re}{\Delta t}}\frac{k_x^2}{2K} p_w
     + \dfrac{ik_x}{K\Delta t} \widehat{\wallquant{\pdv{u}{y}}}\right).
     \label{eq:statond}
\end{equation}

\noindent 
Since the contribution from each term of the original cost function
\eqref{eq:cost} is linear, the weighted sum of each contribution $\phi_i$ as follows:

\begin{equation}
\widehat{\phi}(k_x,k_z)=\sum_{i=1}^{13}a_i\widehat{\phi_i}(k_x,k_z),~\mathrm{or}~\phi(x,z)=\sum_{i=1}^{13}a_i\phi_i(x,z).
\label{eq:control_sum}
\end{equation}
\noindent
Readers are referred to the Appendix for the specific form of $\phi_i$.

The control inputs are updated every time interval of $\Delta t^+= 1.0$ which corresponds to fifty computational timesteps in the present simulations. 
While it is possible to update the control inputs every time step, we followed Choi and Sung \cite{choi_assessment_2002}, which showed that the maximum drag reduction effect could be achieved when the control is updated at specific intervals. 

According to our preliminary numerical test and also existing studies
\cite{han_active_2020,park_machine-learning-based_2020,lee_turbulence_2023}, feedback control with wall measurements can be unstable.
To avoid high-wavenumber oscillations due to the numerical instability, we apply a Gaussian filter $F(K)=\exp(-(\lambda_c/2\pi)^2 K^2)$ to the obtained suboptimal control input derived. 
We minimize the filtering effect by setting the cutoff wavelength $\lambda_c$ to be the relatively small size as $\lambda_c^+=21.9$ in viscous scale.  
As will be shown in Table \ref{tab:drag_reduction_rate}, the present results show essentially the same drag reduction rates as those reported in previous studies for existing control algorithms, and large drag reduction rates are also obtained for new control algorithms obtained from the present Bayesian optimizations.

%% file: BO.tex
\section{Bayesian Optimization and performance indices}
\label{BO}

As described in the previous section, the present cost function contains 10 parameters, i.e., $w_i~(i = 1,2,4-6,8,9,11-13)$. 
They must be optimized to minimize skin friction drag.
The performance of control based on this cost function is anticipated to depend significantly on the selection of these parameters. 
However, finding an appropriate set of parameters through researchers' trial and error is a cumbersome task. 
Therefore, in this study, we introduce Bayesian optimization, a type of black-box optimization method, to explore these coefficients.

Bayesian optimization is a method that constructs a probability-based surrogate model of a black-box system through exploration and performs optimization using this surrogate model. 
Due to its characteristics, it is capable of finding parameter sets that maximize performance far more efficiently compared to grid or random search.
In Bayesian optimization, typically, as shown in Algorithm \ref{algo:bo}, the relationship between design variables $x_n$ and design objectives $y_n$ is expressed using a probability model $S$. 
The next exploration point $x_{n+1}$ is then chosen as the point where the performance is expected to be maximized based on this probability model.

\begin{algorithm}[H]
    \caption{Basic flow of the Bayesian optimization\cite{shahriari_taking_2016,bergstra_algorithms_2011}}\label{algo:bo}
    \hspace*{\algorithmicindent} \textbf{Input} Number of maximum iterations $N$\\
    \hspace*{\algorithmicindent} \textbf{Output} The best design parameter  $x_{\mathrm{optimal}}$
    \begin{algorithmic}[1]
        \For{$n = 0,1,2,\cdots,N$ }
        \State Select new $x_{n+1}$ by optimizing  $x_{n+1}$ = $\mathrm{arg~max}_{x}~EI(x;S_{n})$
        \State Evaluate objective function with simulation to obtain $y_{n+1}=\alpha(x_{n+1})$
        \State Update surrogate model to fit the newly obtained data
        \EndFor
        \State $x_{\mathrm{optimal}}=x_k,~\mathrm{where}~k={\mathrm{arg~max}_{n} y_n}$
    \end{algorithmic}
\end{algorithm}

The metric to be minimized is the time-averaged $C_f\equiv{\tau_w^\dagger}/{(1/2\rho^\dagger U_b^{\dagger,2})}$ value within $t^+\in[600,1200]$ ($t^+=0$; the onset of the control).
$\tau_w^\dagger$ is a spatially averaged streamwise shear stress on the wall.
It is worth noting that the average time period is deliberately kept short to allow for the exploration of a wide range of possibilities, while also minimizing computational expenses during multiple attempts.
Therefore, once a decent algorithm is obtained, we test the algorithm for a longer time, namely $t^+\in[0,10000]$.
In addition to that, if the computation diverges, the $C_f$ is set to be one, which is greater than the typical value of $C_f$ at the Reynolds number, for a penalty.

In Bayesian optimization, as the surrogate model is stochastic, it does not directly try to minimize the value of $C_f$. 
Instead, a separate criterion is established to strike a balance between exploration and exploitation, proposing search points that optimize this criterion. 
In this study, we optimized the following criterion \eqref{eq:ei} known as Expected Improvement. 
\begin{equation}
    \label{eq:ei}
    EI(x;S) = \int_{-\infty}^{\infty} \mathrm{max}(y^*-y,0)S(y|x)dy.
\end{equation}
Here, $S(y|x)$ represents the probability that the objective function value is $y$ at a search point $x$, and $y^*$ is a chosen threshold.

\begin{figure}[h]
    \centering
    \includegraphics[width=0.8\hsize]{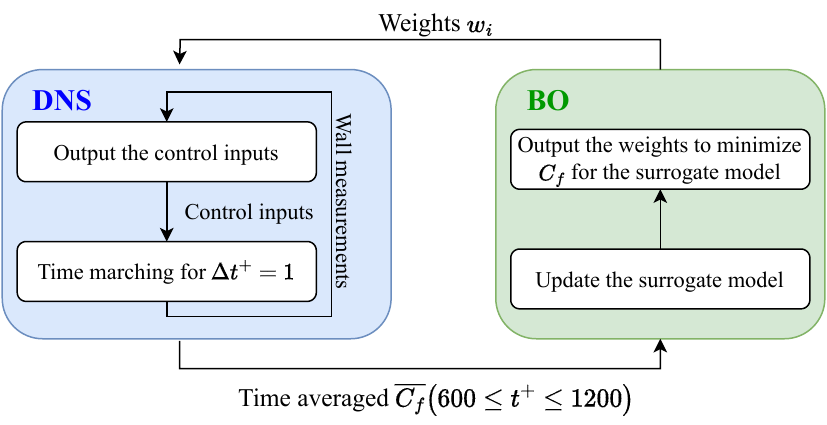}
    \caption{Flowchart of the present optimization framework. The left blue box is a process of direct numerical simulation, whereas the right green box is a process of Bayesian optimization.}
    \label{fig:bo-process}
\end{figure}

The diagram in Fig. \ref{fig:bo-process} outlines the steps involved in the current framework. 
To evaluate a proposed cost function, we conduct a direct numerical simulation of the flow for a duration of $t^+=1200$ with the constant weights $w_i$ output by the Bayesian optimization.
Throughout the optimization, we conducted twenty trials for Case 1 and 150 trials for Case 2 to attain the optimal weights.
The runs were terminated once we confirm that the optimizations had converged.

In the present study, the Bayesian optimization framework is implemented via the open-source Python library Optuna \cite{akiba_optuna_2019} with Tree-structured Parzen Estimator \cite{bergstra_algorithms_2011} as a surrogate functional and fANOVA importance factor \cite{hutter_efficient_2014} as a hyperparameter evaluator.
Readers are referred to the literature for details.

%% file: Results.tex
\section{Results and Discussions}
\label{results}

\subsection{Verification and baseline}

We first verify the present code by comparing the drag reduction rates for the suboptimal control with those proposed in previous studies \cite{lee_suboptimal_1998,choi_assessment_2002,iwamoto_reynolds_2002}. 
Hereinafter, we refer to the cost functions $\mathcal{J}=\left\langle a\left.{\partial u}/{\partial y}\right|_w^2+\phi^2\right\rangle$, $ \mathcal{J}=\left\langle\left.-a{\partial w}/{\partial y}\right|_w^2+\phi^2\right\rangle$ and $\mathcal{J}=\left\langle\left.-a{\partial p}/{\partial z}\right|_w^2+\phi^2\right\rangle$ as LKC-1, LKC-2, and LKC-3, respectively, where $a$ is a positive real number.
The drag reduction rates are averaged over $t^+\in[3000,10000]$ to remove the effect of initial transient, and they are summarized in Tab. \ref{tab:drag_reduction_rate}. 
We also provide reference results from the previous studies \cite{lee_suboptimal_1998,choi_assessment_2002,iwamoto_reynolds_2002}.
The drag reduction rate is defined as $100\times(1-\overline{C_f}/\overline{C_{f,\mathrm{unctrl}}})$, where $\overline{C_f}$ and $\overline{C_{f,\mathrm{unctrl}}}$ indicate the time averaged $C_f$ values for the controlled case and uncontrolled cases, respectively.
Although the present drag reduction rates are slightly lower than the reference for LKC-3, the results generally agree with those of previous studies. 
The small discrepancy can be attributed to the application of the Gaussian filter, the use of a finer grid than that in the reference \cite{lee_suboptimal_1998}, and the strong dependency of the drag reduction rate, especially for LKC-3, on parameters found in Choi and Sung \cite{choi_assessment_2002}.
Furthermore, we have verified that the control inputs shown in Fig. \ref{fig:a} align qualitatively well with Ref. \cite{nakashima_assessment_2017,lee_turbulence_2023}, where both of the control inputs share the streaky structures.

\begin{table}[h]
\centering
\caption{Drag reduction rates of suboptimal control with different cost functions averaged over $t^+\in[3000,10000]$ and their comparison Refs. \cite{lee_suboptimal_1998,choi_assessment_2002,iwamoto_reynolds_2002}. The constant $a$ is a positive real number.}
{
\begin{tabular}{|c|c|c|c|c|}
\hline
cost functions                     & The present study & Lee et al., \cite{lee_suboptimal_1998} & Choi et al., \cite{choi_assessment_2002} & Iwamoto et al., \cite{iwamoto_reynolds_2002} \\ \hline
$\displaystyle \mathcal{J}=\left\langle a\left.\pdv{u}{y}\right|_w^2+\phi^2\right\rangle$,~LKC-1 & -1.7\%            & No effects         & -           & -                      \\ \hline
$\displaystyle \mathcal{J}=\left\langle\left.-a\pdv{w}{y}\right|_w^2+\phi^2\right\rangle$,~LKC-2 & 23.2\%             & 25\%               & 26\% (optimum)                 & about~20\%             \\ \hline
$\displaystyle \mathcal{J}=\left\langle-a\left.\pdv{p}{z}\right|_w^2+\phi^2\right\rangle$,~LKC-3 & 9.2\%             & 16\%               & 12\% (optimum)       & -                      \\ \hline
The best weights in Case 1                 & 22.4\%           & -                  & -                    & -             \\ \hline
The best weights in Case 2                 & 21.6\%           & -                  & -                    & -             \\ \hline
Cf. Opposition control \cite{choi_active_1994}                   & 20.4\%           & -                  & -                    & about~20\%             \\ \hline
\end{tabular}}
\label{tab:drag_reduction_rate}
\end{table}

\subsection{Case 1}

\begin{figure}[h]
    \centering
    \includegraphics[width=0.5\hsize]{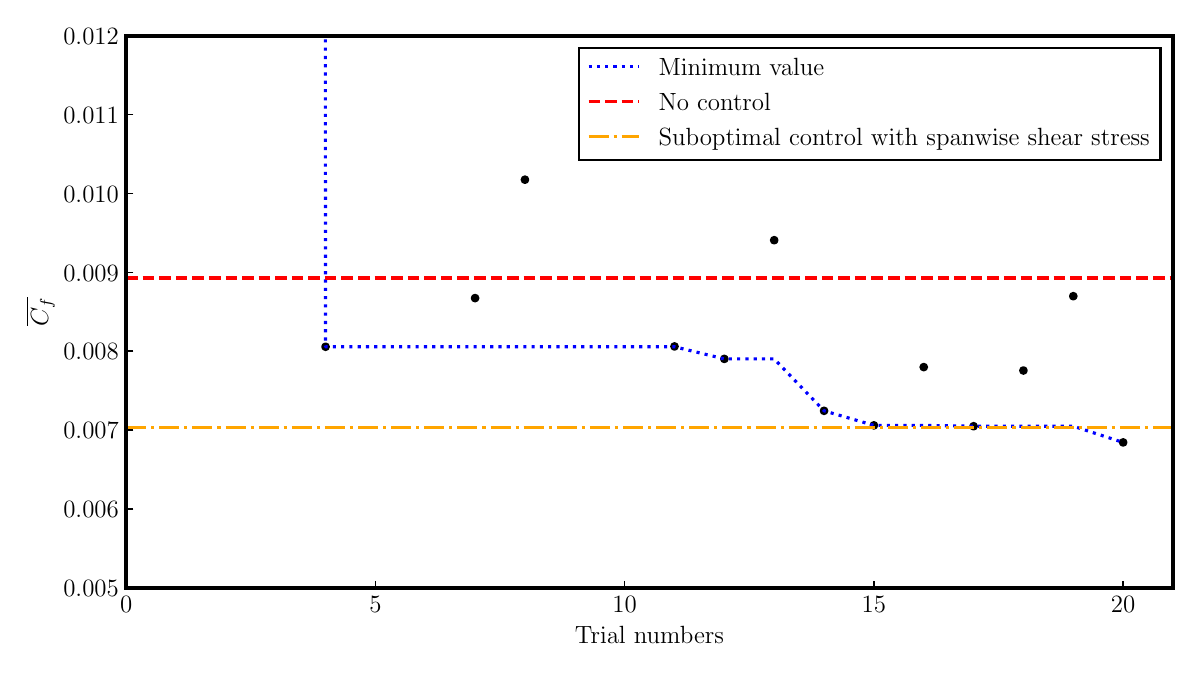}
    \caption{Time averaged friction coefficient $\overline{C_f}$ from $t^+=600$ to $1200$ in the optimization process of Case 1; each black dots indicates each trials; blue dotted line indicates the elite case so far; red dash line indicates uncontrolled case; yellow dash-dot line indicates LKC-2.}
    \label{fig:trials_pilot}
\end{figure}

Here, we first consider Case 1, where the two constants $a_1$ and $a_6$ are optimized in the cost function \eqref{eq:cost_Case 1}.
Again, we note that under the current setup, $a_1=0,a_6<0$ corresponds to LKC-2 and $a_1>0,a_6=0$ corresponds to LKC-1.
Figure \ref{fig:trials_pilot} shows the change of $\overline{C_f}$ with the trial number.
The red dashed line represents the time-averaged $C_f$ of the uncontrolled case, while the yellow line refers to that of LKC-2, which exhibits the highest drag reduction rate among the proposed cost functions.
Although some cases show an increase in drag or divergence, we eventually reach a state where drag is reduced more than the original suboptimal control, at least within the evaluation time range.

The total number of iterations in Case 1 is twenty.
It is generally hard to determine if the black-box optimization is converged. 
Hence we stopped our iterations once we achieved a better drag reduction than the algorithm proposed by Lee et al. \cite{lee_suboptimal_1998}.
Note that Case 1 is considered to validate the present optimization framework for the cost function with only the two design variables.
Therefore, we conducted no further investigations to seek a better result.

\begin{figure}[h]
    \centering
    \includegraphics[width=0.5\hsize]{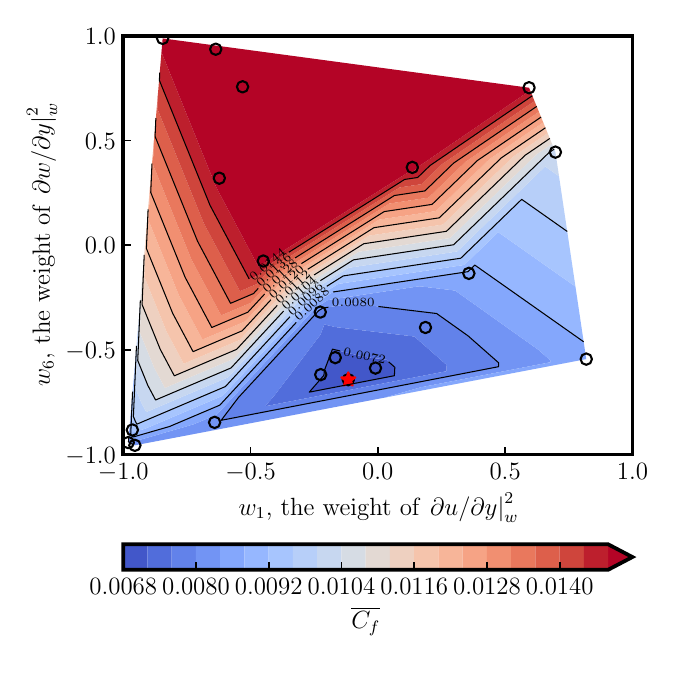}
    \caption{Time averaged $\overline{C_f}$ from $t^+=600$ to $1200$ distribution as a functional of $w_1$ and $w_6$; Red indicates drag-increasing cases while blue indicates drag-reducing cases; circles indicates each trials.; red star indicates the optimum case.}
    \label{fig:param_contour_pilot}
\end{figure}

The contour in Fig. \ref{fig:param_contour_pilot} displays the correlation between $\overline{C_f}$ and the parameters for Case 1, i.e., $w_1$ and $w_6$. 
The performance largely depends on the sign of $w_6$, and it is advantageous for $w_1$ to be approximately zero. 
The trend is somewhat unclear due to the limited number of measurements and short integration time.
As a result, the optimal cost is $(w_1,w_6)=(-0.12,-0.64)$.
Namely, the optimal cost function determined in Case 1 is roughly consistent with LKC-2. 
Therefore, the slight decrease in performance is attributed to the contribution from the streamwise shear stress term.

\begin{figure}[h]
    \centering
    \includegraphics[width=\hsize]{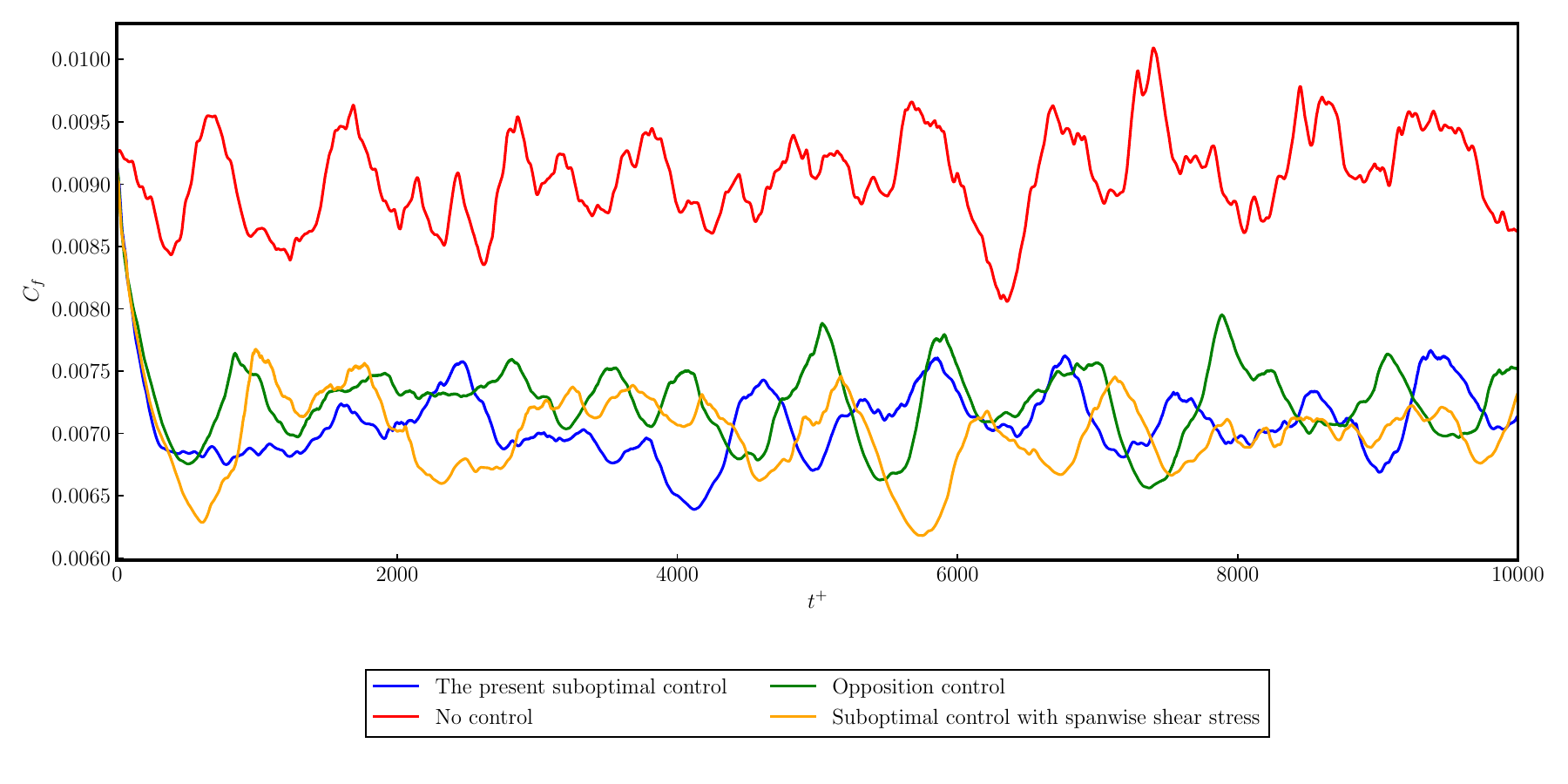}
    \caption{Temporal evolution of $\overline{C_f}$ from the onset of the control until $t^+=10000$ with the best weights in Case 1; red line indicates uncontrolled; blue line indicates the present suboptimal control (Case 1); green line indicates opposition control \cite{choi_active_1994}; yellow line indicates LKC-2.}
    \label{fig:history_pilot}
\end{figure}

Figure \ref{fig:history_pilot} compares the temporal evolution of $C_f$ using the optimized weights from Case 1 with opposition control \cite{choi_active_1994} and LKC-2. 
Between $t^+\in[600,1200]$, the algorithms were more effective than LKC-2 as shown in Fig. \ref{fig:trials_pilot}. 
However, there was no significant improvement beyond LKC-2. 
In fact, according to Table \ref{tab:drag_reduction_rate}, the present algorithm is slightly worse than LKC-2. 

\subsection{Case 2}

\begin{figure}[h]
    \centering
    \includegraphics[width=0.8\hsize]{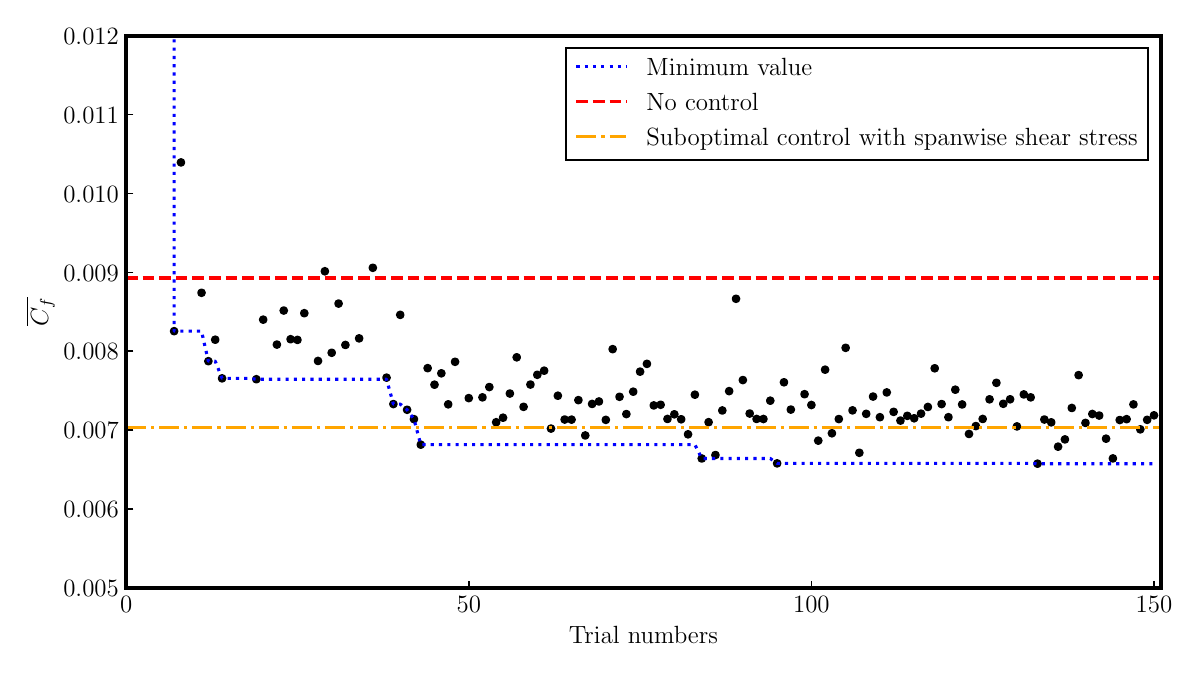}
    \caption{Time averaged friction coefficient $\overline{C_f}$ from $t^+=600$ to $1200$ in the optimization process of Case 2; each black dots indicates each trials; blue dotted line indicates the elite case so far; red dash line indicates uncontrolled case; 
    yellow dash-dot line indicates LKC-2.}
    \label{fig:trials}
\end{figure}

Here, we consider the extended cost function \eqref{eq:cost_Case 2} with ten design variables.
Figure \ref{fig:trials} shows the obtained $C_f$ with the trial number. 
Although about 50 trials have yielded drag reduction rates that exceed that of LKC-2, the trial-and-error process was continued to ensure that the optimization was converged.
After 150 trials, we obtain the lowest value of $C_f$ around $6.5\times10^{-3}$, slightly lower than $C_f$ achieved with the existing cost functions.

\begin{figure}[h]
\centering
\begin{minipage}[b]{0.49\columnwidth}
    \centering
    \includegraphics[width=1.0\columnwidth]{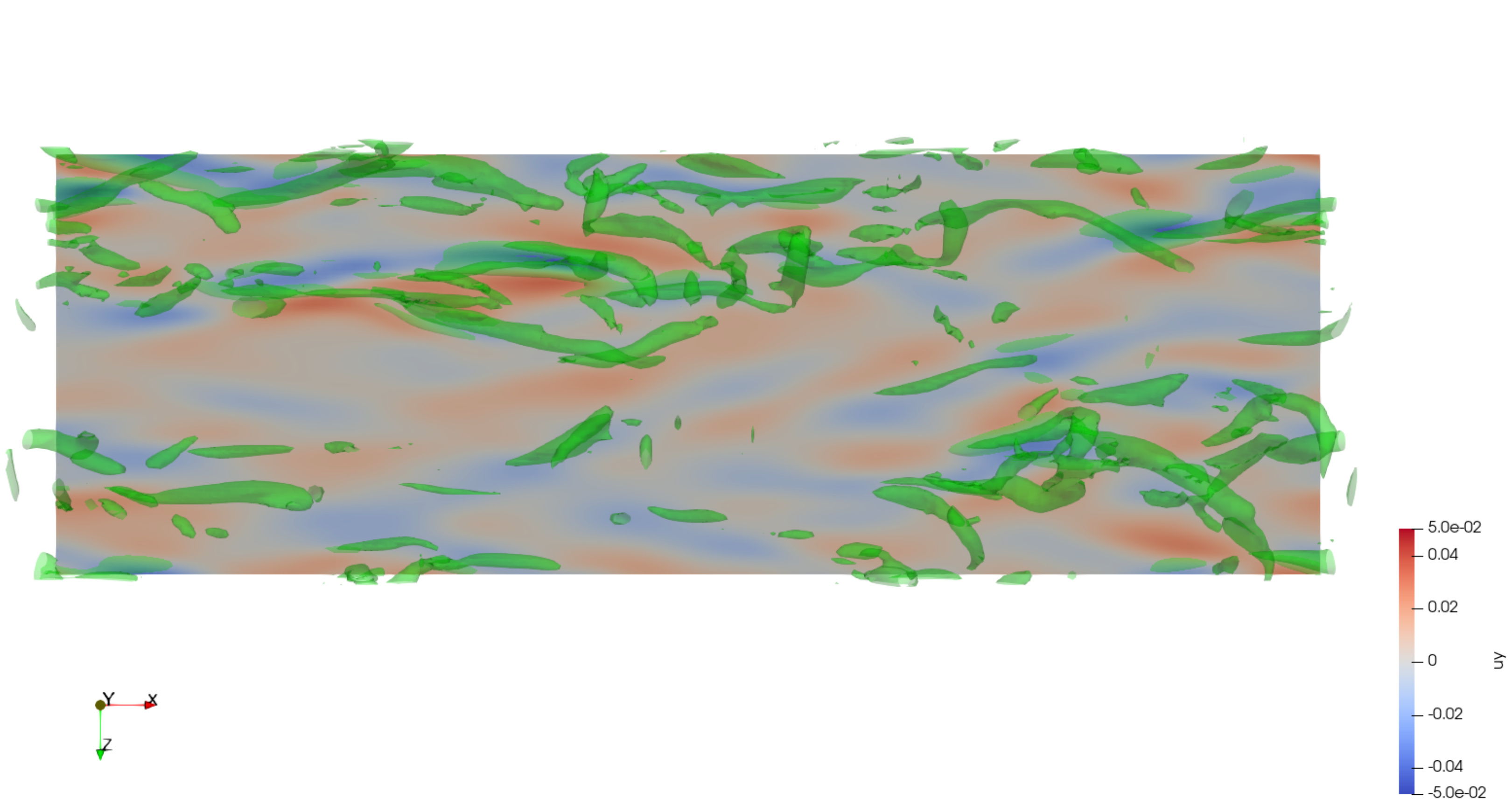}
    \subcaption{LKC-2}
    \label{fig:a}
\end{minipage}
\begin{minipage}[b]{0.49\columnwidth}
    \centering
    \includegraphics[width=1.0\columnwidth]{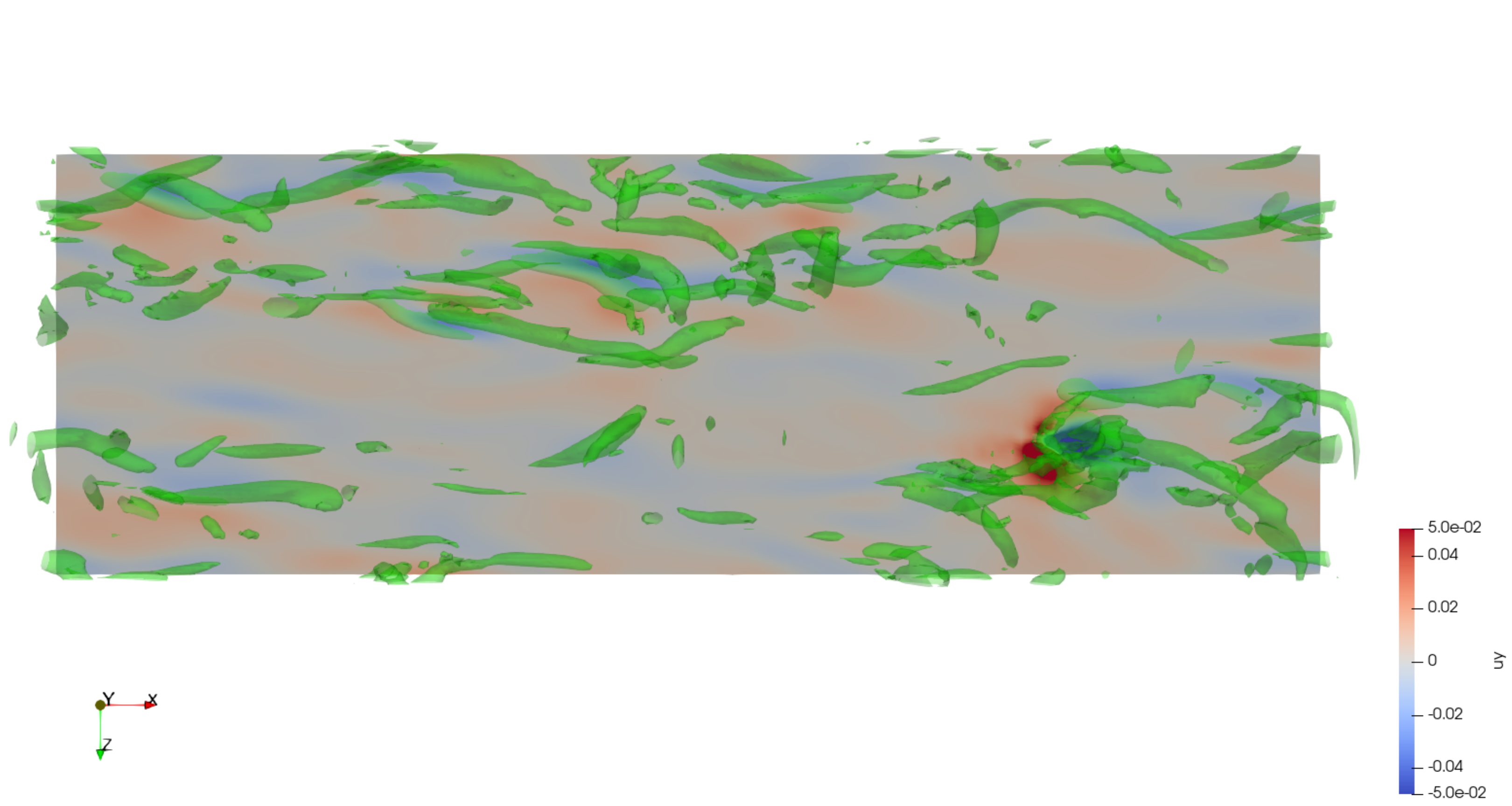}
    \subcaption{Case 2}
    \label{fig:b}
\end{minipage}
\caption{Control inputs and vortical structures around the onset of the control.
The red color shows blowing, while the blue color shows suction on the wall.
The green three-dimensional isosurface is the surface of $Q^+=0.012$, where $Q$ is the second invariant of the velocity gradient tensor.}
\end{figure}

In Figs. \ref{fig:a} and \ref{fig:b}, the control inputs resulting from both LKC-2 and the best cost function obtained in the present study are shown with the vortical structures visualized by the second invariant of velocity gradient tensor.
Although both have a similar streak-like structure, the control law obtained in the present study indicates that strong suction and blowing are applied at the location where the vortex structure is concentrated.

\begin{figure}[h]
    \centering
    \includegraphics[width=\hsize]{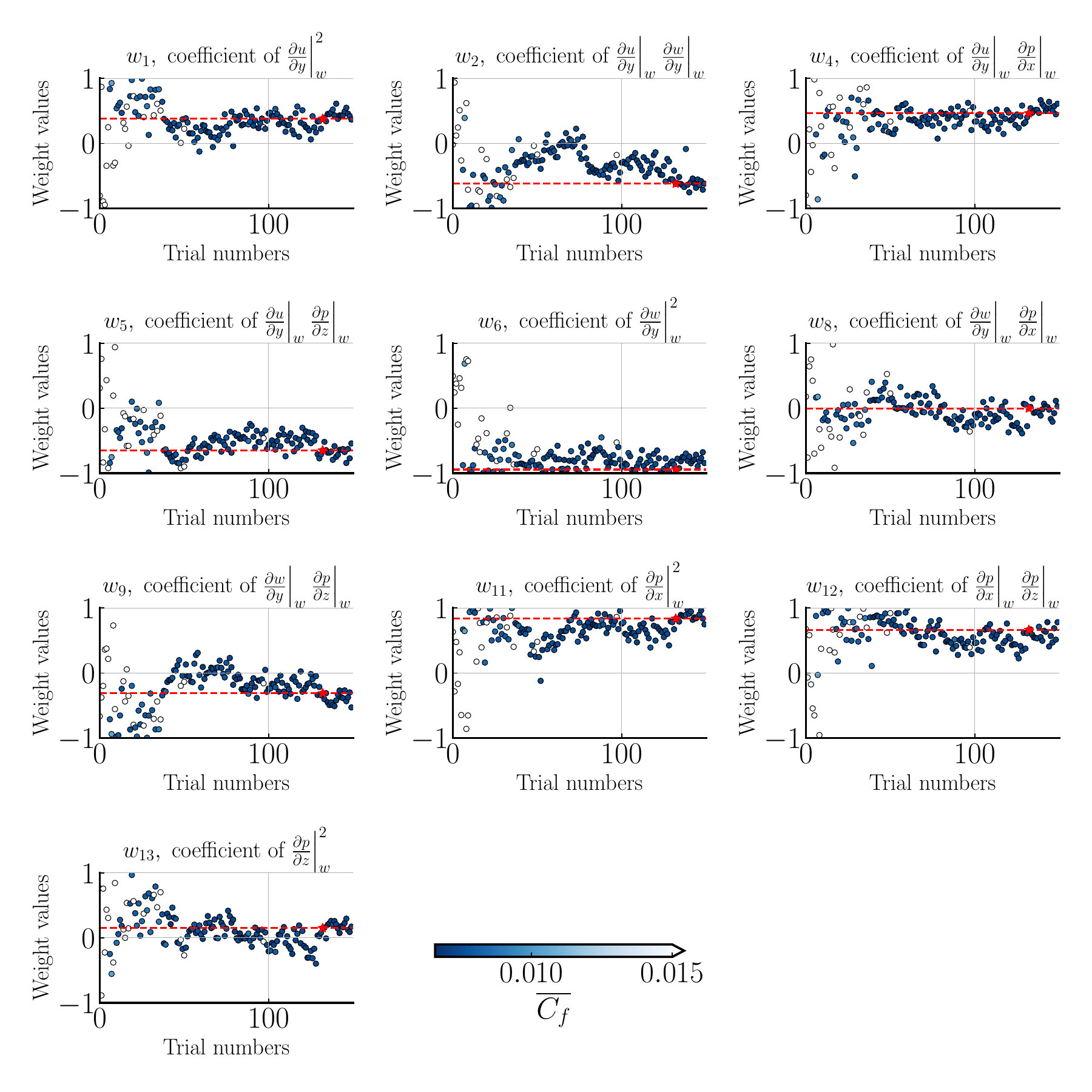}
    \caption{Weights parameter distribution as a function of trial number; circle indicates each trial; circle color indicates their metrics. Thick blue represents drag-reduction while white represents drag-increase; the red star indicates the best case; the red dash line indicates the best case weight values.}
    \label{fig:param_dist}
\end{figure}

Table \ref{tab:optimal_weights} presents the best parameters obtained throughout the present optimization, and Fig. \ref{fig:param_dist} provides the convergence processes of each weighting coefficient.
In Fig. \ref{fig:param_dist}, the darker color is representing lower drag, and vice versa.
For $w_6$ (the weight of $\displaystyle\qty({\partial w}/{\partial y})^2$), a clear trend is observed. 
The optimal pattern is predominantly at the lower limit (-1), and when $w_6<0$, the drag reduction effect is high, while it is low when $w_6>0$.
Moreover, for $w_1$ (the weight of $\displaystyle\qty({\partial u}/{\partial y})^2$), patterns with a drag reduction effect are scattered around zero. 
This observation aligns with Lee et al. \cite{lee_suboptimal_1998}, confirming that including the fluctuation intensity of the streamwise shear stress in the cost function has a limited effect.

Non-zero weights are assigned to other parameters, suggesting that they have a non-negligible impact on the control effect. 
For example, significant positive values are assigned to $w_{11}$ and $w_{12}$ for $\displaystyle\qty({\partial p}/{\partial x})^2$ and $\displaystyle{\partial p}/{\partial x}\cdot{\partial p}/{\partial z}$.
These results may be related to feedback control based on vorticity flux, as mentioned by Koumoutsakos \cite{koumoutsakos_vorticity_1999}. 
Additionally, despite the effectiveness of including $\displaystyle\qty({\partial p}/{\partial z})^2$ in the cost function as confirmed in Lee et al. \cite{lee_suboptimal_1998} and Choi and Sung \cite{choi_assessment_2002} and also in the present Case 1, the optimal weight coefficient for this term , i.e. $w_{13}$, is relatively small.

\begin{table}[h]
\centering
\caption{The optimal coefficients obtained from the present optimization. $w_{\{3,7,10\}}$ are manually set zero.}
\label{tab:optimal_weights}
\begin{tabular}{c|c|c|c|c|c|c}
    $w_1$  & $w_2$  & $w_3$  & $w_4$  & $w_5$  & $w_6$  & $w_7$ \\ \hline
    0.23  & -0.31 & 0     & 0.43  & -0.52 & -0.85 & 0    \\ \hline\hline
    $w_8$  & $w_9$  & $w_{10}$ & $w_{11}$ & $w_{12}$ & $w_{13}$ &      \\ \hline
    -0.29 & -0.11 & 0     & 0.79  & 0.40  & 0.11  &      \\ 
\end{tabular}
\end{table}

\begin{figure}[h]
    \centering
    \includegraphics[width=0.5\hsize]{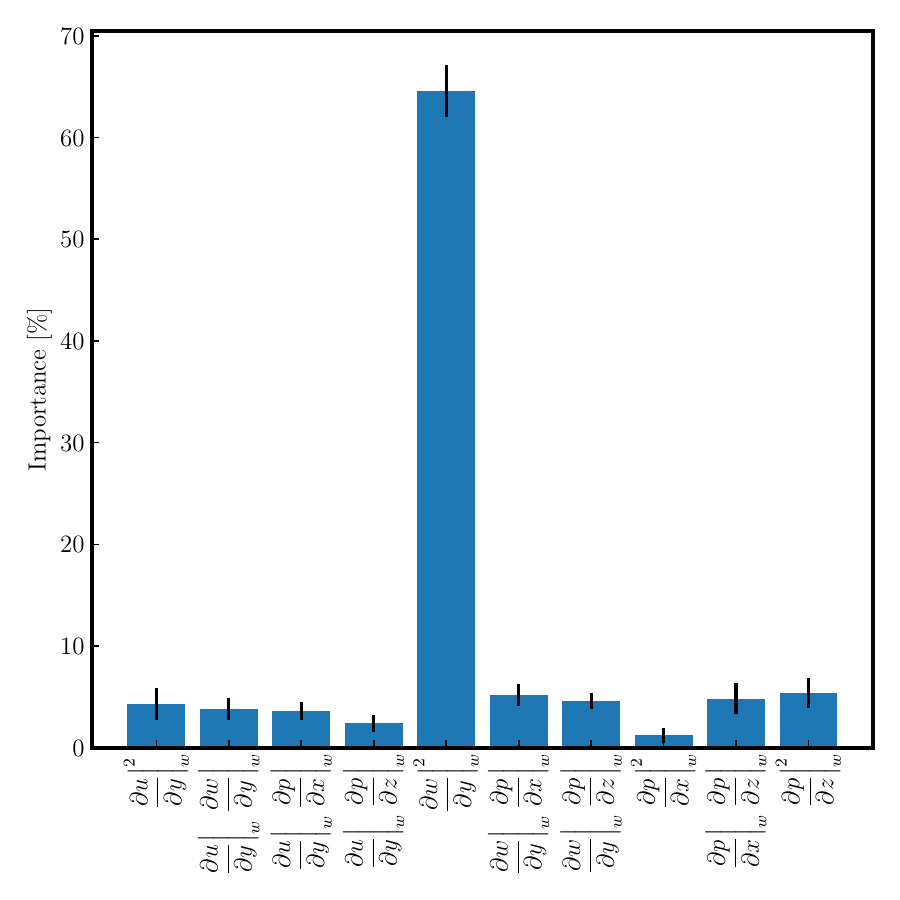}
    \caption{fANOVA importance \cite{hutter_efficient_2014} among the present weights to be optimized in Case 2; The values show the importance given to the weight parameters with the metrics.; The vertical black lines indicate their standard deviation through ten computations.}
    \label{fig:fanova_importance}
\end{figure}

To analyze the importance of each parameter, we introduce the fANOVA framework \cite{hutter_efficient_2014}. 
The framework uses a random forest tree to predict the sensitivity of each parameter, with a larger value indicating greater sensitivity. 
We computed the importance based on fifty randomly sampled trials to ensure reliable results and trained the random forest tree ten times. 
The results are summarized in Fig. \ref{fig:fanova_importance}, revealing that $\displaystyle({\partial w}/{\partial y})^2$ was the only parameter that had significant importance among ten variables considered in Eq. \eqref{eq:cost_Case 2}. 
This result aligns with our observation in Fig. \ref{fig:param_dist} that the distribution of parameters, excluding $w_6$, did not appear sensitive to the performance.
Our results may be interpreted as proof of optimality in the choice of cost function by Lee et al. \cite{lee_suboptimal_1998} within the framework of suboptimal control, even though our exploration space is still limited to the prescribed library.

\begin{figure}[h]
    \centering
    \includegraphics[width=\hsize]{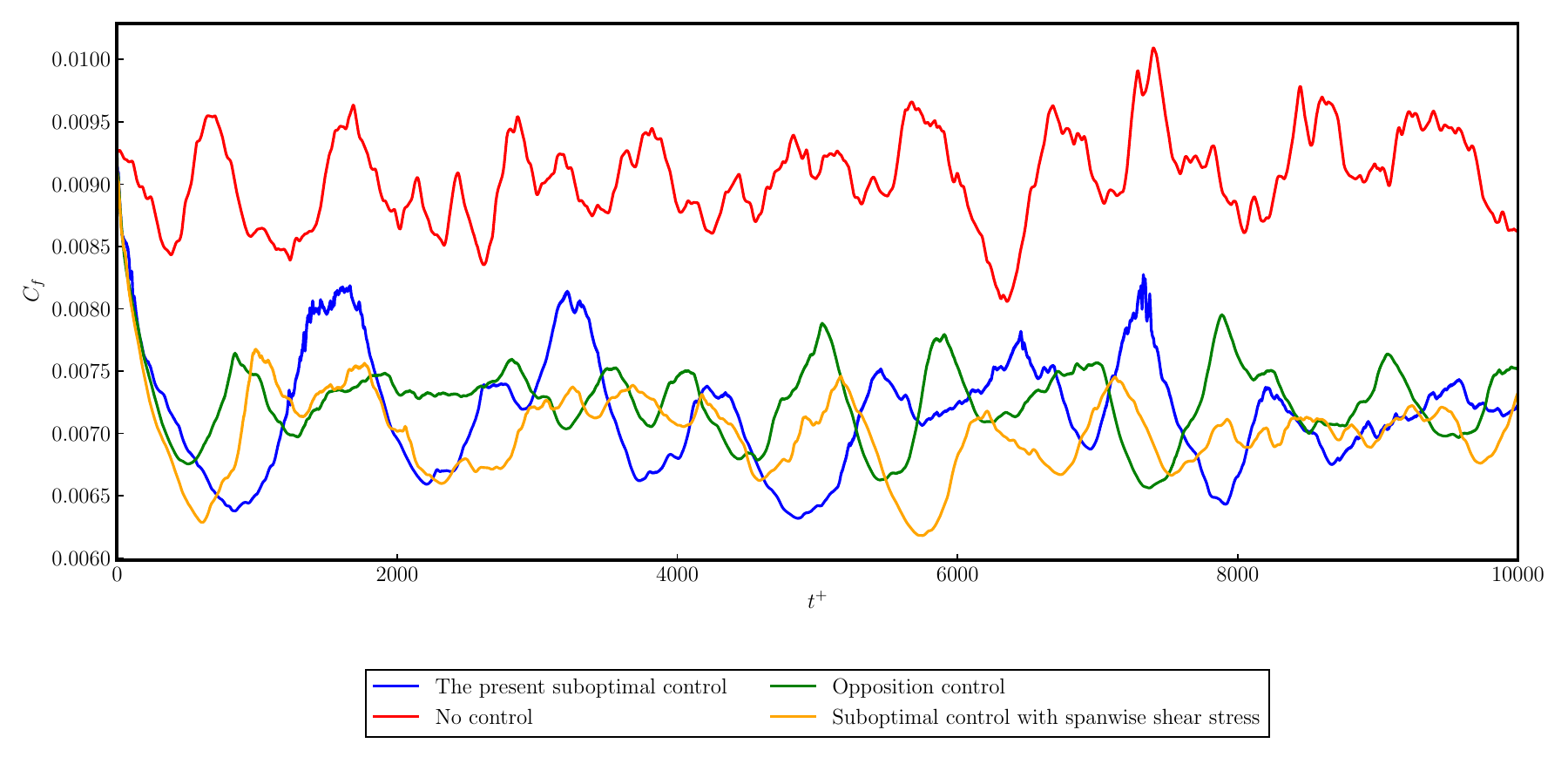}
    \caption{Temporal evolution of $\overline{C_f}$ from the onset of the control until $t^+=10000$ with the best weights in Case 2; red line indicates uncontrolled; blue line indicates the present suboptimal control (Case 2); green line indicates opposition control \cite{choi_active_1994}; yellow line indicates the original suboptimal control \cite{lee_suboptimal_1998} with spanwise shear stress.}
    \label{fig:history}
\end{figure}

Since the relatively short integration time of $T^+=1200$ is used during the optimization to reduce the computational cost,
we evaluate the obtained control algorithms with a longer integration time of $t^+\in[0,10000]$ after the optimization.
The results are summarized in Fig. \ref{fig:history} and Tab. \ref{tab:drag_reduction_rate}.
While a higher drag reduction effect than LKC-2 is observed within the short integration time used in the optimization process, the results of longer simulations reveal that the high drag reduction effect could not be confirmed.
Consequently, as summarized in Tab. \ref{tab:drag_reduction_rate}, the control performance is $21.6\%$, slightly lower than that of LKC-2 and even Case 1, although we consider the cost function with larger degrees of freedom in Case 2.
In addition, the time history of $C_f$ obtained in Case 2 shows a larger time scale variation $(T^+\sim O(10^3))$ than that of Case 1 and the original algorithm.
In the present optimization setting, the time average of a limited interval $t^+\in[600,1200]$ was employed. 
This problem could be mitigated by employing a more extended evaluation period during the optimization.

%% file: Conclusion.tex
\section{Conclusion}
\label{conclusion}

In this study, we proposed a method using Bayesian optimization to develop a novel feedback control law via the optimization of cost function for suboptimal control. 
We considered two cost functions: one incorporating the fluctuations of streamwise and spanwise shear stress (Case 1), and the other incorporating the quadratic of wall shear stresses and pressure gradients (Case 2). 
The former comprises two terms, while the latter includes ten terms.
We determined the weight coefficients for each term using Bayesian optimization, naturally extending the cost function proposed by Lee et al. (1998) \cite{lee_suboptimal_1998}. 
By applying Bayesian optimization, we found weights that yield control results equivalent to existing suboptimal control by Lee et al. \cite{lee_suboptimal_1998} for both cases.
Our results revealed that including the fluctuation intensity of the spanwise wall shear stress in the cost function is effective for drag reduction in both cases. 
This observation aligns well with the results reported by Lee et al. \cite{lee_suboptimal_1998}. 
From here, we verified that it is feasible to find a cost function for optimal control of turbulent flow through an automatic manner. 
Although our study focused on controlling turbulent flow in a channel, the proposed method is applicable to other problem settings. 

Finally, we note some open questions regarding the present work.
Though we employed a library-based approach to optimize the cost function, a library-free approach (e.g., genetic programming \cite{koza_genetic_1994}) is another option and may overcome the potential limitation of the present approach.
It is also crucial to evaluate the effectiveness of the proposed method in various applications of inverse problems within thermo-fluid systems.
Finally, in the present framework, the coefficients in the cost function do not have to be kept constant during the evaluation period of a cost function.
Such flexible adjustment of the variables in the cost function shall also be considered in future studies.

%% file: Appendix.tex
\section{Specific form of $\phi_i$}

Each $\phi_i$ is derived as:
\begin{equation}
  \widehat{\phi_1}= -\sqrt{\dfrac{2\re}{\Delta t}}\dfrac{ik_x}{K}{\displaystyle\left.\pdv{\widehat{u}}{y}\right|_w},
  \label{eq:phi1}
\end{equation}
\begin{equation}
  \widehat{\phi_2} = -\sqrt{\dfrac{2\re}{\Delta t}}\left(\displaystyle\dfrac{ik_z}{2K}\left.\pdv{\widehat{u}}{y}\right|_w+\displaystyle\dfrac{ik_x}{2K}\left.\pdv{\widehat{w}}{y}\right|_w\right),
  \label{eq:phi2}
\end{equation}
\begin{equation}
    \widehat{\phi_3}=-\left(\dfrac{1}{K\Delta t}\left.\pdv{\widehat{u}}{y}\right|_w+\dfrac{ik_x}{2K}\sqrt{\dfrac{2\re}{\Delta t}}\widehat{p_w}\right),
    \label{eq:phi3}
\end{equation}
\begin{equation}
  \widehat{\phi_4}= \left(\displaystyle\dfrac{k_x^2}{2K}\sqrt{\dfrac{2\re}{\Delta t}}\widehat{p_w}+\dfrac{ik_x}{K\Delta t}\left.\pdv{\widehat{u}}{y}\right|_w\right),
  \label{eq:phi4}
\end{equation}
\begin{equation}
  \widehat{\phi_5}=  \left(\displaystyle\dfrac{k_xk_z}{2K}\sqrt{\dfrac{2\re}{\Delta t}}\widehat{p_w}+\dfrac{ik_z}{K\Delta t}\left.\pdv{\widehat{u}}{y}\right|_w\right),
  \label{eq:phi5}
\end{equation}
\begin{equation}
  \widehat{\phi_6}= -\sqrt{\dfrac{2\re}{\Delta t}}\dfrac{ik_z}{K}{\displaystyle\left.\pdv{\widehat{w}}{y}\right|_w},
  \label{eq:phi6}
\end{equation}
\begin{equation}
    \widehat{\phi_7}=-\left(\dfrac{1}{K\Delta t}\left.\pdv{\widehat{w}}{y}\right|_w+\dfrac{ik_z}{2K}\sqrt{\dfrac{2\re}{\Delta t}}\widehat{p_w}\right)
    \label{eq:phi7}
\end{equation}
\begin{equation}
  \widehat{\phi_8}=  \left(\displaystyle\dfrac{k_xk_z}{2K}\sqrt{\dfrac{2\re}{\Delta t}}\widehat{p_w}+\dfrac{ik_x}{K\Delta t}\left.\pdv{\widehat{w}}{y}\right|_w\right),
  \label{eq:phi8}
\end{equation}
\begin{equation}
  \widehat{\phi_9}= \left(\displaystyle\dfrac{k_z^2}{2K}\sqrt{\dfrac{2\re}{\Delta t}}\widehat{p_w}+\dfrac{ik_z}{K\Delta t}\left.\pdv{\widehat{w}}{y}\right|_w\right),
  \label{eq:phi9}
\end{equation}
\begin{equation}
    \widehat{\phi_{10}} = -\dfrac{2}{K\Delta t}\widehat{p_w}
    \label{eq:phi10}
\end{equation}
\begin{equation}
  \widehat{\phi_{11}}= -\dfrac{2k_x^2}{K\Delta t}\widehat{p_w},
  \label{eq:phi11}
\end{equation}
\begin{equation}
  \widehat{\phi_{12}} = -\dfrac{2k_xk_z}{K\Delta t}\widehat{p_w},
  \label{eq:phi12}
\end{equation}
\begin{equation}
  \widehat{\phi_{13}}= -\dfrac{2k_z^2}{K\Delta t}\widehat{p_w}.
  \label{eq:phi13}
\end{equation}
    
It should be noted that the initial component of Eq. \eqref{eq:phi2} and the subsequent component of Eq. \eqref{eq:phi5} has a similar equational form as the successful control strategy suggested by previous studies \cite{kawagoe_proposal_2019}. 
The former was obtained using genetic algorithms, while the latter was proposed through resolvent analysis. 
Both resulted in a 10\% reduction in drag.
In both cases, however, the control inputs were generated without considering the governing equations (in the former case) or the suboptimal control theory (in the latter case), while the present study derives the above expression based on the suboptimal control theory.